\documentclass[12pt,preprint]{aastex}

\slugcomment{  }

\shorttitle{Postbounce evolution of core-collapse supernovae}
\shortauthors{Sumiyoshi et al.}

\begin{document}


\title{Postbounce evolution of core-collapse supernovae:\\
       Long-term effects of equation of state}


\author{K. Sumiyoshi
}
\affil{Numazu College of Technology, \\
       Ooka 3600, Numazu, Shizuoka 410-8501, Japan}
\email{sumi@numazu-ct.ac.jp}

\author{S. Yamada
}
\affil{Science and Engineering, Waseda University, \\
       Okubo, 3-4-1, Shinjuku, Tokyo 169-8555, Japan \\
       \& \\
       Advanced Research Institute for Science and Engineering, Waseda University, \\
       Okubo, 3-4-1, Shinjuku, Tokyo 169-8555, Japan}

\author{H. Suzuki
}
\affil{Faculty of Science and Technology, Tokyo University of Science, \\
       Yamazaki 2641, Noda, Chiba 278-8510, Japan}

\author{H. Shen
}
\affil{Department of Physics, Nankai University, \\
       Tianjing 300071, China}

\author{S. Chiba
}
\affil{Advanced Science Research Center, \\
       Japan Atomic Energy Research Institute, \\
       Tokai, Ibaraki 319-1195, Japan}

\and

\author{H. Toki
}
\affil{Research Center for Nuclear Physics (RCNP), Osaka University, \\
       Mihogaoka 10-1, Ibaraki, Osaka 567-0047, Japan}




\begin{abstract}
We study the evolution of supernova core 
from the beginning of gravitational collapse 
of a 15M$_\sun$ star up to 1 second after core bounce.  
We present results of spherically symmetric simulations of 
core-collapse supernovae 
by solving general relativistic $\nu$-radiation-hydrodynamics 
in the implicit time-differencing.  
We aim to explore the evolution of shock wave in a long term 
and investigate the formation of protoneutron star together with 
supernova neutrino signatures.
These studies are done 
to examine the influence of equation of state (EOS) 
on the postbounce evolution of shock wave in the late phase 
and the resulting thermal evolution of protoneutron star.  
We make a comparison of 
two sets of EOS, that is, by Lattimer and Swesty (LS-EOS) and by Shen et al.(SH-EOS).  
We found that, for both EOSs, the core does not explode and 
the shock wave stalls similarly in the first 100 milliseconds 
after bounce.  
The revival of shock wave does not occur even after a long period 
in either cases.  
However, the recession of shock wave appears 
different beyond 200 milliseconds after bounce, having different 
thermal evolution of central core.  
A more compact protoneutron star is found for LS-EOS than SH-EOS 
with a difference in the central density by a factor of $\sim$2  
and a difference of $\sim$10 MeV in the peak temperature.  
Resulting spectra of supernova neutrinos are different 
to the extent that may be detectable by terrestrial neutrino detectors.

%
%
%

\end{abstract}



\keywords{supernovae: general --- stars: neutron --- 
neutrinos --- hydrodynamics --- equation of state}


\section{Introduction}

Understanding the explosion mechanism of core-collapse supernovae 
is a grand challenge that requires endeavor to conduct numerical 
simulations of $\nu$-radiation-hydrodynamics with best knowledge of 
particle and nuclear physics.  
Three dimensional simulations of 
$\nu$-radiation-hydrodynamics, which are currently formidable, and 
better determinations of  the nuclear equation of state of dense matter 
and the neutrino-related reaction rates are mandatory.  
One has to advance step by step by developing numerical methods 
and examining microphysics and its influence in various stages 
of supernovae.  

Even with the extensive studies in recent years 
with currently available computing resources, 
the numerical results have not made clear the explosion mechanism.  
On one hand, recent multi-dimensional supernova simulations 
with approximate neutrino-transport schemes 
have revealed the importance of asymmetry such as rotation, convection, 
magnetic fields and/or hydrodynamical instability \citep{blo03,bur03,kot03,fry04,kot04,wal04}.  
On the other hand, 
spherically symmetric supernova simulations of late 
have removed the uncertainty of neutrino-transport and 
clarified the role of neutrinos in core-collapse and 
shock propagation \citep{ram00,lie01,mez01,ram02,tho03}.  
In this study, we focus on spherically symmetric simulations 
which are advantageous to examine the role of microphysics 
without ambiguity of neutrino-transport.  

Almost all authors have reported that 
neither prompt explosion nor delayed explosion occurs 
under the spherical symmetry.  
This conclusion is commonly reached by simulations 
with Newtonian \citep{ram00phd,ram00,mez01,tho03}, 
approximately relativistic \citep{ram02} and 
fully general relativistic \citep{lie01} gravity, 
together with standard microphysics, i.e. the equation of state (EOS) 
by \citet{lat91} and the weak reaction rates by \citet{bru86}.  
The influence of nuclear physics inputs has been 
further assessed by employing the extended 
neutrino reactions \citep[][see also section \ref{sec:weak}]{tho03} and 
more up-to-date electron capture rates 
on nuclei \citep{hix03}.  
The dependence on the progenitor models \citep{lie02,tho03,lie04,jan04} 
and the sets of physical EOS \citep{jan04} has been studied very recently.  
These simulations so far have shown that the collapse of 
iron cores leads to the stalled shock after bounce without 
successful explosion.  

In the current study, we explore the influence of EOS 
in the time period that has not been studied very well in the previous 
studies.  
Most of recent numerical simulations have been performed 
until about 300 milliseconds after bounce.  
This is due to the severe limitation on time steps by the 
Courant condition for explicit time-differencing schemes.  
A typical time step in the explicit method is about 
10$^{-6}$ seconds after the formation of dense compact objects.  
However, in the implicit method as we employ in this study, 
the time step is not restricted by the Courant condition.  
This is advantageous for a long-term evolution.  
In the studies by \citet{lie02,lie04}, who also adopted the implicit 
method, the postbounce evolution 
has been followed up to about 1 second for a small number of models 
with Lattimer and Swesty EOS.  
Historically, the idea of the delayed explosion was 
proposed by Wilson's simulations that followed more than 
several hundred milliseconds after bounce.
In some cases, the revival of shock wave occurred even 
beyond 0.5 seconds \citep[for example,][]{bet85}.  
It is still interesting to explore this late phase 
in the light of possible influence of microphysics.  

The progress of the supernova EOS put an additional 
motivation to the study of this late postbounce phase.  
Only recently the sets of physical EOS, which cover 
a wide range of density, composition and temperature 
in a usable and complete form, have become 
available for simulations.  
A table of EOS was made for the first time by 
\citet{hil85} within 
the Skyrme Hartree-Fock approach and 
applied to some simulations \citep{hil84,suz90,suz93,suz94,sum95c,jan04}.  
Another set of EOS has been provided as a numerical routine by 
\citet{lat91} utilizing the compressible 
liquid-drop model.  
This EOS has been used these years as a standard.  

Recently, a new complete set of EOS 
for supernova simulations has become available \citep{she98a,she98b}.  
The relativistic mean field (RMF) theory with a local density approximation 
was applied to the derivation of the table of supernova EOS.  
This EOS is different in two important aspects from previous 
EOSs.  
One thing is that the Shen's EOS is based on the relativistic 
nuclear many-body framework whereas the previous ones are based 
on the non-relativistic frameworks.  
The relativistic treatment is known to affect the behavior 
of EOS at high densities (i.e. stiffness) \citep{bro90}
and the size of nuclear symmetry energy \citep{sum95b}.  
The other thing is that the Shen's EOS is based on the experimental 
data of unstable nuclei, which have become available recently.  
The data of neutron-rich nuclei, which are close to the 
astrophysical environment, were used to constrain the 
nuclear interaction.  
The resulting properties of isovector interaction are generally 
different from the non-relativistic counterpart and the 
size of symmetry energy is different.  
The significant differences in stiffness and compositions 
during collapse and bounce have been shown 
between Shen's EOS and Lattimer-Swesty EOS 
by hydrodynamical calculations \citep{sum04}.  
Therefore, it would be exciting to explore the supernova dynamics 
with the new set of EOS.  
Such an attempt has been made recently by \citet{jan04} and 
no explosion has been reported up to 300 milliseconds 
after bounce.  

Our aim of the current study is, therefore, the comparison 
of the postbounce evolutions beyond 300 milliseconds for the first time.  
We perform the core-collapse simulations adopting the two sets of 
EOS, that is, Shen's EOS (SH-EOS) and Lattimer-Swesty EOS (LS-EOS).  
We follow the evolutions 
of supernova core for a long period.  
We explore the fate of the stalled shock 
up to 1 second after bounce.  
In this time period, one can also see the birth of protoneutron star 
as a continuous evolution from the collapsing phase 
together with the long-term evolution of neutrino emissions.
Although the supernova core does not display successful explosion, as 
we will see, the current simulations may provide some aspects of 
central core leading to the formation of 
protoneutron star or black hole.  
This information is also helpful to envisage the properties of 
supernova neutrinos in the first second since the simulations of 
protoneutron star cooling done so far 
usually starts from several hundred milliseconds 
after bounce for some given profiles.  
As a whole, 
we aim to clarify how the EOS influences the dynamics of shock wave, 
evolution of central core and supernova neutrinos.  

\section{Numerical Methods}

A new numerical code of general relativistic, $\nu$-radiation-hydrodynamics 
under the spherical symmetry has been developed 
\citep{yam97,yam99} for supernova simulations.  
The code solves a set of equations of hydrodynamics and 
neutrino-transfer simultaneously in the implicit way, 
which enables us to have substantially longer time steps than 
explicit methods.  
This is advantageous for the study of long-term behaviors 
after core bounce.
The implicit method has been also adopted by \citet{lie04}
in their general relativistic $\nu$-radiation-hydrodynamics code.  
They have taken, however, an operator splitting method 
so that hydrodynamics and neutrino-transfer could be treated separately.  

\subsection{Hydrodynamics}
The equations of lagrangian hydrodynamics 
in general relativity are solved by a implicit, 
finite differencing.  
The numerical method is based on the approximate 
linearized Riemann solver (Roe-type scheme) that 
captures shock waves without introducing 
artificial viscosities.  
Assuming the spherical symmetry, the metric 
of \citet{mis64} is adopted to formulate hydrodynamics 
and neutrino-transport equations.  
A set of equations for the conservation of baryon number, 
lepton number and energy-momentum are solved together 
with the metric equations and the entropy equation.  
Details of the numerical method of hydrodynamics 
can be found in \citep{yam97}, where standard numerical 
tests of the hydrodynamics code have been also reported.  

\subsection{Neutrino-transport}
The Boltzmann equation for neutrinos in general relativity 
is solved by a finite difference scheme (S$_N$ method) implicitly 
together with above-mentioned lagrangian hydrodynamics.  
The neutrino distribution function, $f_\nu (t, m, \mu, \varepsilon_\nu)$, 
as a function of time $t$, lagrangian mass coordinate $m$, neutrino 
propagation angle $\mu$ and neutrino energy $\varepsilon_\nu$, is evolved.  
Finite differencing of the Boltzmann equation is mostly based 
on the scheme by \citet{mez93a}.  
However, the update of time step 
is done simultaneously with hydrodynamics.  
The reactions of neutrinos are explicitly 
calculated in the collision terms of the Boltzmann equation 
with incident/outgoing neutrino angles and energies taken into account.  
Detailed comparisons with the Monte Carlo method have been 
made to validate the Boltzmann solver and 
to examine the angular resolution \citep{yam99}.  

\subsection{$\nu$-radiation-hydrodynamics}
The whole set of finite-differenced equations described above are solved by 
the Newton-Raphson iterative method.
The Jacobian matrix forms a block-tridiagonal matrix, in which 
dense block matrices arise from the collision terms of the transport 
equation.  
Since the inversion of this large matrix is most costly in the 
computing time, 
we utilize a parallel algorithm of block cyclic reduction 
for the matrix solver \citep{sum98}.
In the current simulations, we adopt non-uniform 255 spatial zones 
for lagrangian mass coordinate.
We discretize the neutrino distribution function 
with 6 angle zones and 14 energy zones 
for $\nu_e$, $\bar{\nu}_e$, $\nu_{\mu/\tau}$ and $\bar{\nu}_{\mu/\tau}$, respectively.

\subsection{Rezoning}\label{sec:rezoning}
The description of long-term evolution of accretion 
in a lagrangian coordinate is a numerically tough problem.   
In order to keep enough resolution during the accretion phase, 
rezoning of accreting materials is 
done long before they accrete onto the 
surface of protoneutron star and become opaque to neutrinos.  
At the same time, dezoning of the hydrostatic inner part of 
protoneutron star is done 
to avoid the increase of grid points.  

When we have tried simulations without rezoning, neutrino luminosities 
oscillate largely in time due to intermittent accretion of coarse grid points 
and it sometimes leads to erroneous dynamics (even explosions).  
Therefore, we have checked that the resolution of grid points is enough 
by refining the initial grid points and rezoning during the simulations.  
Even then, there are still slight oscillations in luminosities and average 
energies of neutrinos in the last stage of calculations.  
There are also transient kinks sometimes when the grid size in mass 
coordinate changes 
during accretion as we will see in section \ref{sec:snnu}.  
These slight modulations of neutrino quantities, however, do not affect 
the overall evolution of protoneutron stars with accretion 
once we have enough resolution.  

\section{Model Descriptions}

As an initial model, we adopt the profile of iron core of 
a 15M$_{\odot}$ progenitor from \citet{woo95}.  
This progenitor has been widely used in supernova simulations.  
The computational grid points in mass coordinate are non-uniformly 
placed to cover the central core, shock propagation region and 
accreting material with enough resolution.  

\subsection{Equation of state}
The new complete set of EOS for supernova simulations (SH-EOS) 
\citep{she98a,she98b} is derived by 
the relativistic mean field (RMF) theory with a local density approximation.  
The RMF theory has been a successful framework 
to reproduce the saturation properties, masses and radii 
of nuclei, and proton-nucleus scattering data \citep{ser86}.
We stress that the RMF theory \citep{sug94} is based 
on the relativistic Br\"uckner-Hartree-Fock (RBHF) theory 
\citep{bro90}, which is a 
microscopic and relativistic many-body theory.  
The RBHF theory has been shown to be successful to reproduce the 
saturation of nuclear matter starting from the 
nucleon-nucleon interactions determined by 
scattering experiments.  
This is in good contrast with non-relativistic 
many-body frameworks which can account for the saturation 
only with the introduction of extra three-body 
interactions.  

The effective interactions in the RMF theory 
have been determined by least squares fittings to reproduce 
the experimental data of masses and radii of stable and 
unstable nuclei \citep{sug94}.  
The determined parameters of interaction, TM1, have been 
applied to many studies of nuclear structures and experimental 
analyses \citep{sug96,hir97}.  
One of stringent tests on the isovector interaction is passed 
in excellent agreement of the theoretical prediction with the experimental data on 
neutron and proton distributions in isotopes including 
neutron-rich ones with neutron-skins \citep{suz95,oza01}.  
The RMF theory with the parameter set TM1 
provides uniform nuclear matter with 
the incompressibility of 281 MeV 
and the symmetry energy of 36.9 MeV.  
The maximum mass of neutron star is 2.2 M$_{\odot}$ 
for the cold neutron star matter in the RMF with TM1 \citep{sum95a}.  
The table of EOS covers a wide range of density, 
electron fraction and temperature for supernova simulations, 
and has been applied to numerical simulations of 
r-process in neutrino-driven winds \citep{sum00}, 
prompt supernova explosions \citep{sum01}, and 
other simulations \citep{sum95c,ros03,sum04,jan04}.

For comparison, we also adopt the EOS by \citet{lat91}.  
The LS-EOS is based on the compressible liquid drop model for 
nuclei together with dripped nucleons.  
The bulk energy of nuclear matter is expressed in terms 
of density, proton fraction and temperature.  
The values of nuclear parameters are chosen 
according to nuclear mass formulae and other theoretical 
studies with the Skyrme interaction.  
Among various parameters, the symmetry energy is set to be 29.3 MeV, 
which is smaller than the value in the relativistic EOS.  
As for the incompressibility, 
we use 180 MeV, 
which has been used frequently for recent supernova simulations.
In this case, the maximum mass of neutron star is estimated to be 
1.8 M$_{\odot}$.
This choice enables us to make comparisons with previous works, 
though 180 MeV is smaller than the standard value as will be discussed below.  
The sensitivity to the incompressibility of LS-EOS 
has been studied by \citet{tho03} using the choices of 180, 220 and 375 MeV.  
The numerical results of core-collapse and bounce with different 
incompressibilities turn out to be similar up to 200 milliseconds 
after bounce.  
The differences in luminosities and average energies of emergent 
neutrinos are within 10 \% and do not affect significantly the 
post-bounce dynamics on the time scale of 100 ms.  
The influence of different incompressibilities in LS-EOS on 
the time scale of 1 sec remains to be seen as an extension 
of the current study.  
For densities below 10$^7$ g/cm$^3$, the subroutine of 
Lattimer-Swesty EOS runs into numerical troubles, therefore, 
we adopt Shen's EOS in this density regime instead.  
This is mainly for numerical convenience.  
In principle, it is preferable to adopt the EOS, which contains 
electrons and positrons at arbitrary degeneracy and relativity, 
photons, nucleons and an ensemble of 
nuclei as non-relativistic ideal gases 
(see for example, \citet{tim99,tho03}).  
One also has to take into account non-NSE abundances 
determined from the preceding quasi-static evolutions.  
Note that we are chiefly concerned with the effect of EOS at high densities, and 
this pragmatic treatment does not have any significant influence on the shock dynamics.  

We comment here on the nuclear parameters of EOS and its consequences 
for the astrophysical applications considered here.  
The value of incompressibility of nuclear matter has been considered to be 
within 200--300 MeV from experimental data and theoretical analyses.  
The value recently obtained within the non-relativistic approaches \citep{col04} 
is 220--240 MeV.  
The corresponding value extracted within the relativistic approaches 
is known to be higher than non-relativistic counterpart 
and is 250--270 MeV \citep{vre03,col04}.  
It is also known that the determination of incompressibility 
is closely related with the size of the symmetry energy and 
its density dependence.  
The incompressibility of EOS in the RMF with TM1 is 
slightly higher than those standard values 
and the SH-EOS is 
relatively stiff.  
The neutron stars with SH-EOS are, therefore, less compact 
with lower central densities and have higher maximum masses 
than those obtained by LS-EOS with the incompressibility of 180 MeV.  
The adiabatic index of SH-EOS at the bounce of supernova core 
is larger than that of LS-EOS \citep{sum04}.  

The value of symmetry energy at the nuclear matter density 
is known to be around 30 MeV by nuclear mass formulae \citep{mol95}.
The recent derivation of the symmetry energy in a relativistic 
approach gives higher values of 32--36 MeV together with 
the above mentioned higher incompressibility \citep{die03,vre03}.  
The symmetry energy in the RMF with TM1 
is still a bit larger compared with the standard values.  
We note that the symmetry energy in the RMF is determined by the 
fitting of masses and radii of various nuclei 
including neutron-rich ones.  
The large symmetry energy in SH-EOS leads to large 
proton fractions in cold neutron stars, which may lead to 
a possible rapid cooling by the direct URCA process, as well 
as the stiffness of neutron matter \citep{sum95a}.  
The difference between neutron and proton chemical potentials 
is large and leads to different compositions of free protons 
and nuclei \citep{sum04}.  
The consequences of these differences 
in incompressibility and 
symmetry energy will be discussed in the comparison of 
numerical simulations in section \ref{sec:results}.  

\subsection{Weak reaction rates}\label{sec:weak}
The weak interaction rates regarding neutrinos are evaluated 
by following the standard formulation by \citet{bru86}.  
For the collision term in the Boltzmann equation, 
the scattering kernels are explicitly calculated in terms of 
angles and energies of incoming and outgoing neutrinos \citep{mez93b}.  
In addition to the Bruenn's standard neutrino processes, 
the plasmon process \citep{bra93} and 
the nucleon-nucleon bremsstralung process \citep{fri79,max87} are 
included in the collision term.  
The latter reaction has been shown to be an important process 
to determine the supernova neutrinos from the protoneutron star 
cooling \citep{suz93,bur00} as a source of $\nu_{\mu/\tau}$.  
The conventional {\it standard} weak reaction rates are used for the current 
simulations to single out the effect of EOS and to compare 
with previous simulations.  
Recent progress of neutrino opacities in nuclear matter \citep{bur05} 
and electron capture rates on nuclei \citep{lan03} will be examined 
along with the updates of EOS in future studies.  


\section{Comparison of results}\label{sec:results}
We present the results of two numerical simulations 
performed with Shen's EOS and Lattimer-Swesty EOS.  
They are denoted by SH and LS, respectively.  

\subsection{Shock propagation}\label{sec:shock}
\begin{figure}
\epsscale{.60}
\plotone{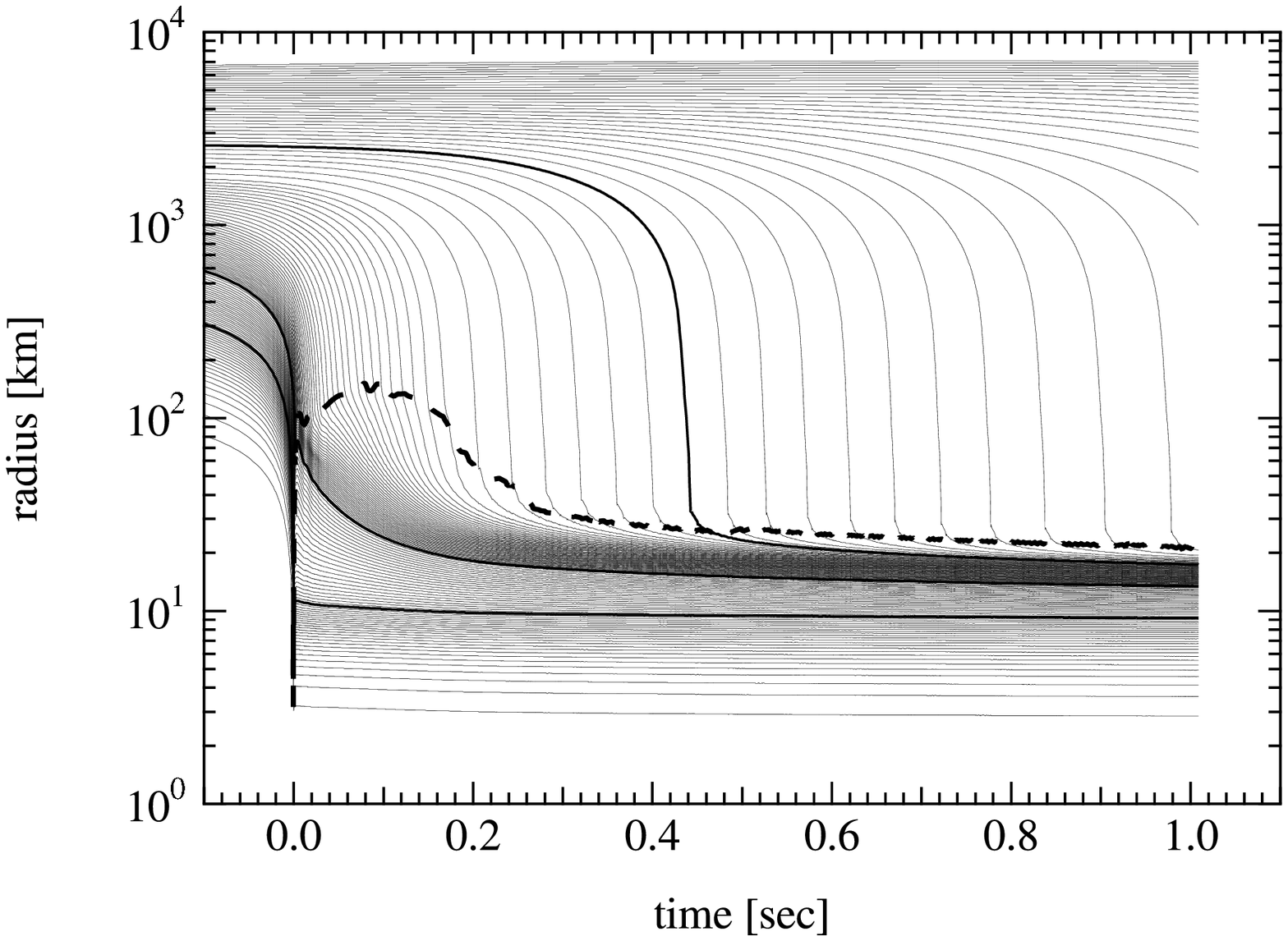}
\caption{Radial trajectories of mass elements of 
the core of 15M$_{\odot}$ star as a function of time after bounce in model SH.
The location of shock wave is displayed by a thick dashed line.}
\label{fig:traj}
\end{figure}

Fig. \ref{fig:traj} shows the radial trajectories of mass elements 
as a function of time after bounce in model SH.  
The trajectories are plotted for each 0.02M$_{\odot}$ in mass coordinate 
up to 1.0M$_{\odot}$ and for each 0.01M$_{\odot}$ for the rest of outer part.
Thick lines denote the trajectories for 0.5M$_{\odot}$, 1.0M$_{\odot}$ and 1.5M$_{\odot}$.
One can see the shock wave is launched up to 150 kilometers 
and stalled there within 100 milliseconds.  
The shock wave recedes down to below 100 kilometers afterwards 
and the revival of shock wave or any sign of it is not found 
even after 300 milliseconds.  

Instead, the stationary accretion shock is formed at several 
tens of kilometers.  
As the central core gradually contracts, a protoneutron star 
is born at center.
The material, which was originally located in the outer core, 
accretes onto the surface of the protoneutron star.  
The accretion rate is about 0.2M$_{\odot}$/s on average 
and decreases 
from 0.25M$_{\odot}$/s to 0.15M$_{\odot}$/s gradually.
This behavior is similar in model LS.  
At 1 sec after bounce, the baryon mass of protoneutron 
star is 1.60M$_{\odot}$ for both cases.  

\begin{figure}
\epsscale{0.35}
\plotone{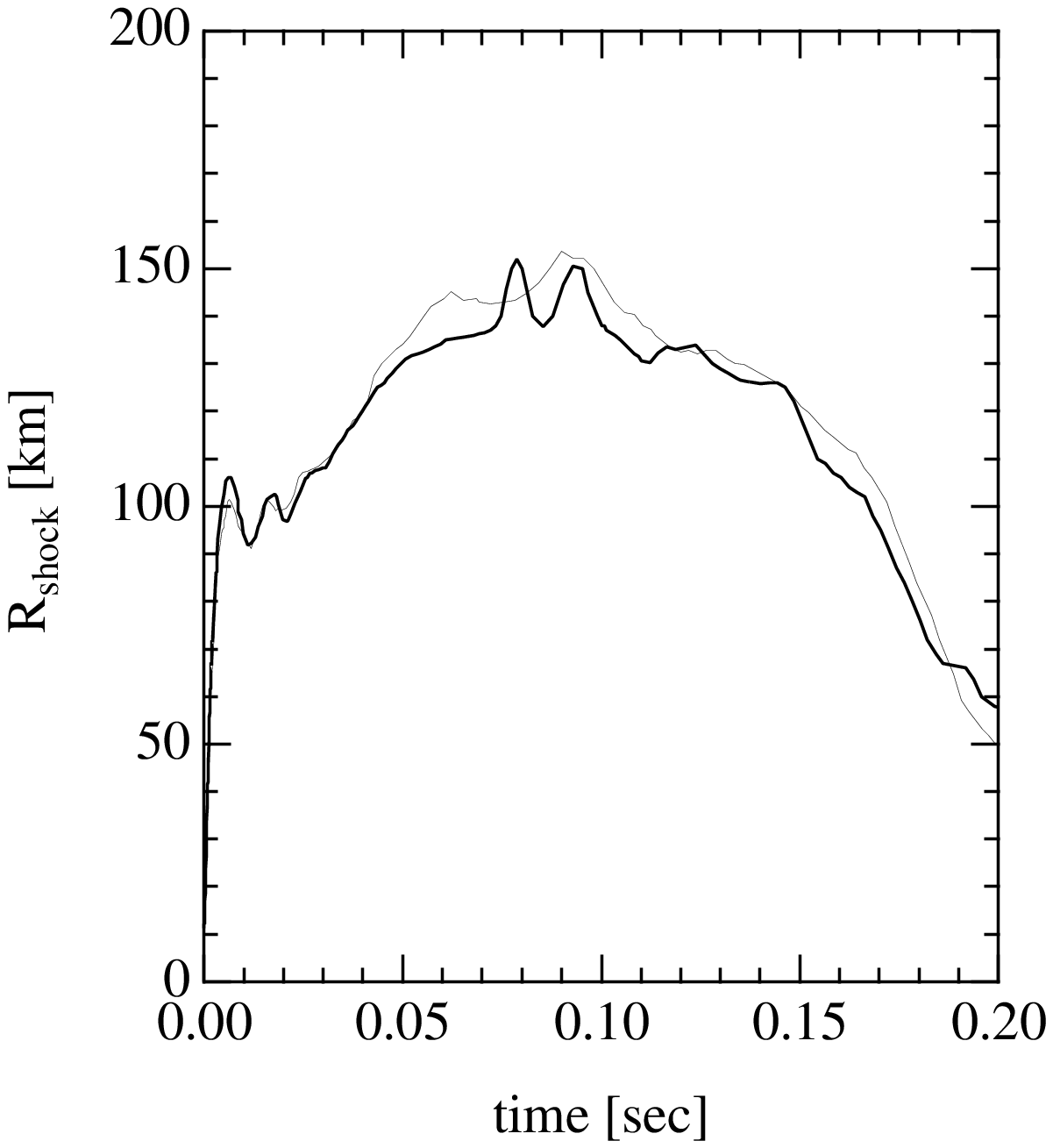}
\epsscale{0.52}
\plotone{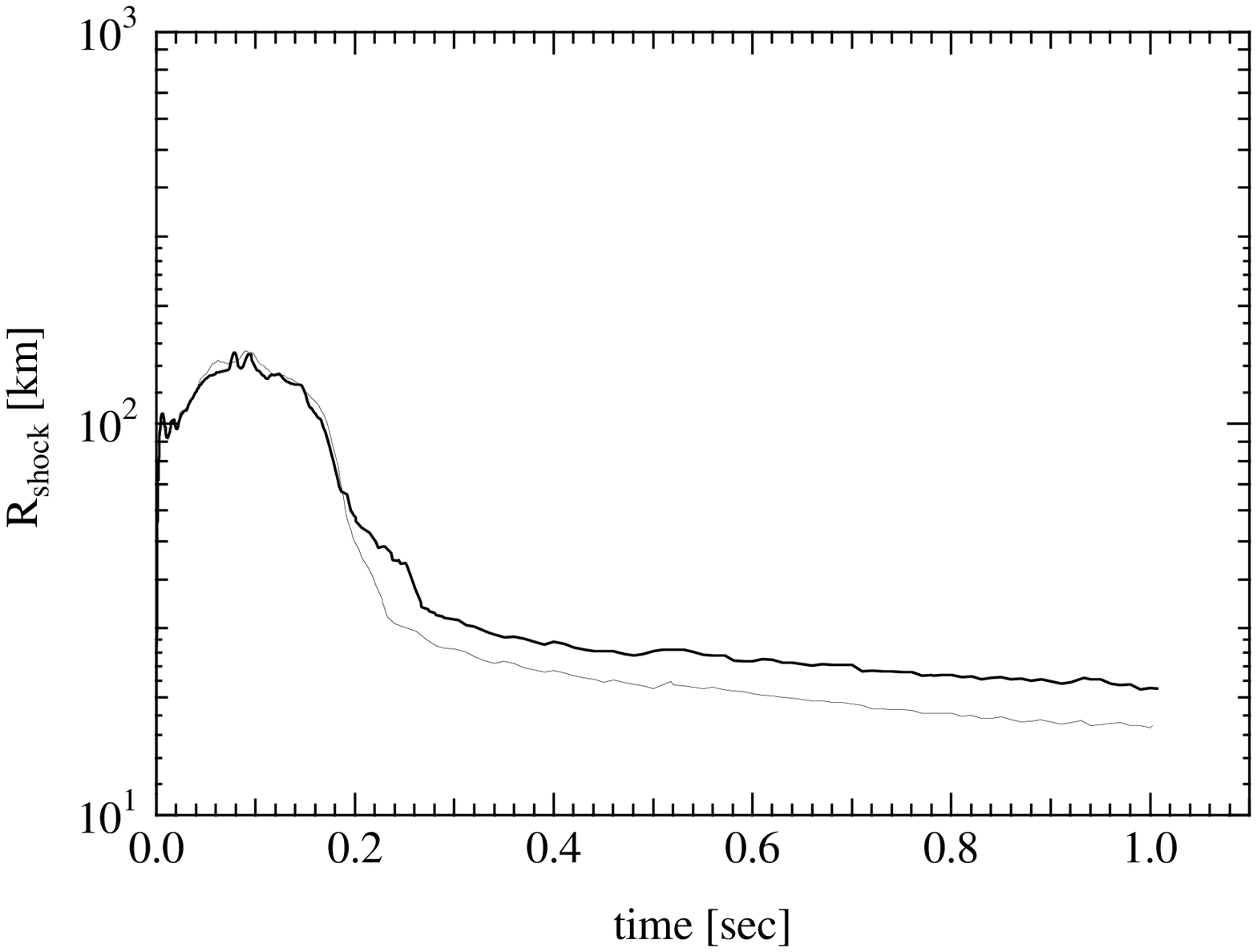}
\caption{Radial positions of shock waves in models SH and LS are 
shown by thick and thin lines, respectively, 
as a function of time after bounce.
The evolutions at early and late times are displayed 
in left and right panels, respectively.
Small fluctuations in the curves are due to numerical artifact 
in the procedure to determine the shock position from a limited 
number of grid points.}
\label{fig:shock}
\end{figure}

The trajectories of shock wave in models SH and LS are 
compared in Fig. \ref{fig:shock}.  
The propagations of shock wave in two models are similar in 
the first 200 milliseconds (left panel).  
We note that slight fluctuations in the curves are 
due to numerical artifact 
in the procedure to determine the shock position.  
Note that we have rather low resolutions in the central part 
in order to have higher resolutions in the accreting material. 
Except for the discrepancy due to the different numerical methods 
(e.g. approximate general relativity, eulerian etc.), zoning and resolutions, 
the current simulations up to 200 milliseconds are consistent 
with the results (middle panel of Figure 3) by \citet{jan04} 
having similar maximum radii and timing of recession.
The difference shows up from 200 milliseconds after bounce 
and becomes more apparent in the later phase (right panel).  
After 600 milliseconds, the shock position in model LS 
is less than 20 kilometers and 
it is clearly different from that in model SH.
This difference originates from the faster contraction of 
the protoneutron star in model LS.  
We discuss the evolution of protoneutron star later 
in section \ref{sec:pns}.  

\subsection{Collapse phase}

The initial propagation of shock wave is largely controlled 
by the properties of the inner core during the gravitational 
collapse.  
We have found noticeable differences in the behavior of 
core-collapse in two models.  
However, they did not change 
the initial shock energy drastically, which then leads to the 
similarity of the early phase of shock propagation we have 
just seen above.

\begin{figure}
\epsscale{.60}
\plotone{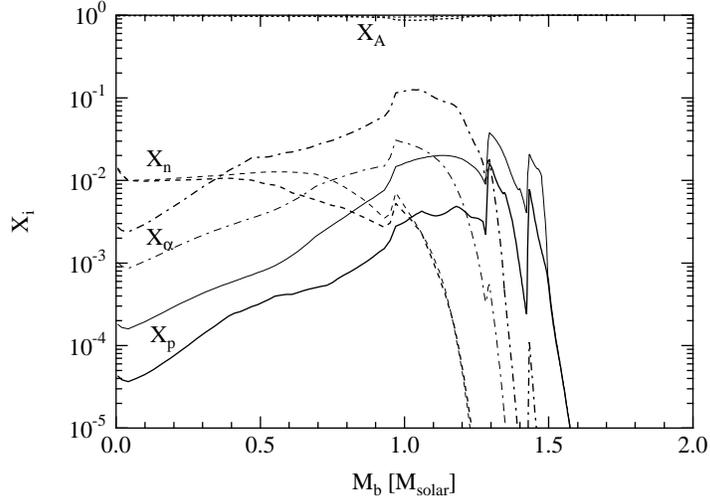}
\caption{Mass fractions in the supernova cores are 
shown as a function of baryon mass coordinate at 
the time when the central density reaches 10$^{11}$ g/cm$^3$.  
Solid, dashed, dotted and dot-dashed lines 
show mass fractions of protons, neutrons, nuclei and alpha 
particles, respectively.
The results for models SH and LS are shown by thick and thin lines, 
respectively.}
\label{fig:xi}
\end{figure}

First of all, it is remarkable that the compositions of 
dense matter during the collapse are different.  
In Fig. \ref{fig:xi}, the mass fraction 
is shown as a function of mass coordinate 
when the central density reaches 10$^{11}$ g/cm$^3$.  
The mass fraction of free proton in model SH is smaller 
than that in model LS by a factor of $\sim$5.  
This is caused by the larger symmetry energy in SH-EOS,
where the proton chemical potential is lower 
than the neutron chemical potential as discussed 
in \citet{sum04}.  
The smaller free proton fraction reduces the electron 
captures on free protons.  
Note that the electron capture 
on nuclei is suppressed in the current simulations 
due to the blocking above N=40 in Bruenn's prescription.  
This is in accordance with the numerical results 
by \citet{bru89,swe94} who studied the influence 
of the free proton fraction and the symmetry energy.
However, there is also a negative feedback in the 
deleptonization during collapse \citep{lie02}.  
Smaller electron capture rates keep electron fraction 
high, which then leads to an increase of free proton fraction 
and consequently to electron captures after all.  
The resultant electron fraction turns out to be not 
significantly different as we will see later.  

\begin{figure}
\epsscale{.60}
\plotone{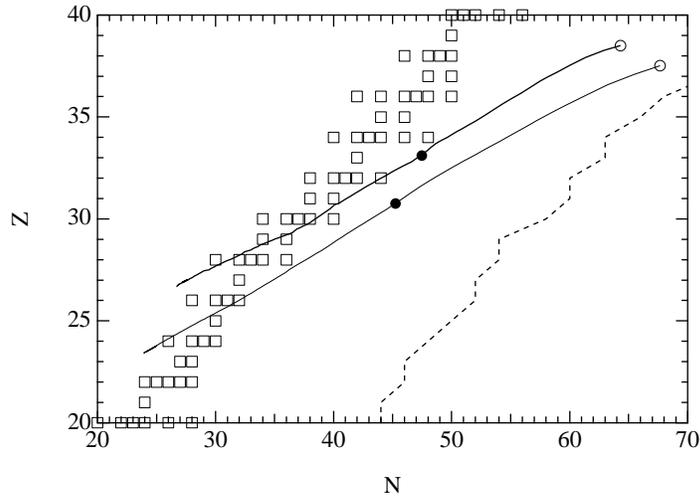}
\caption{Nuclear species appear in supernova cores are shown 
in the nuclear chart.
Stable nuclei and the neutron drip line \citep{hor00} are shown by open square 
symbols and dashed line, respectively.  
Nuclear species at the center of the core are marked by solid circle 
($\rho_c$=10$^{11}$ g/cm$^{3}$) and open circle 
($\rho_c$=10$^{12}$ g/cm$^{3}$) symbols.  
The results for models SH and LS are shown by thick and thin lines, 
respectively.}
\label{fig:nz}
\end{figure}

It is also noticeable that the mass fraction of alpha particles 
differs substantially and the abundance of nuclei is slightly 
reduced in model SH.  
This difference of alpha abundances in two models persists during the 
collapse and even in the post-bounce phase.
The nuclear species appearing in the central core during 
collapse are shown in the nuclear chart (Fig. \ref{fig:nz}).  
The nuclei in model SH are always less neutron-rich than 
those in model LS by more than several neutrons.  
This is also due to the effect of the symmetry energy, which 
gives nuclei closer to the stability line in model SH.  
The mass number reaches up to $\sim$80 and $\sim$100 at the central 
density of 10$^{11}$ g/cm$^3$ (solid circle) and 
10$^{12}$ g/cm$^3$ (open circle), respectively.  
In the current simulations, the electron capture on 
nuclei is suppressed beyond N=40 due to the simple prescription 
employed here 
and the difference of species do not give any difference.
However, results may turn out different when more realistic 
electron capture rates are adopted \citep{hix03}.  
It would be interesting to see whether the difference 
found in two EOSs leads to differences in central cores 
using recent electron capture rates on nuclei \citep{lan03}.  
Further studies are necessary to discuss the abundances 
of nuclei and the influence of more 
updated electron capture rates for mixture of nuclear species 
beyond the approximation of single-species in the current EOSs.  

\begin{figure}
\epsscale{.60}
\plotone{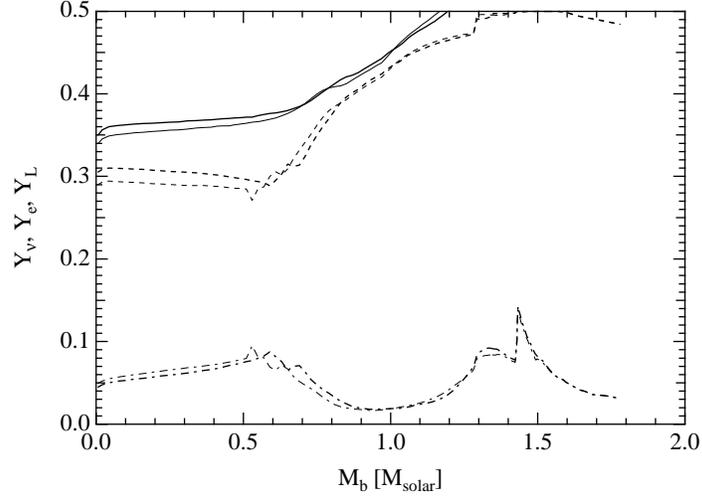}
\caption{Lepton, electron and neutrino fractions at bounce are shown 
as a function of baryon mass coordinate by solid, dashed 
and dot-dashed lines, respectively.
The results for models SH and LS are shown by thick and thin lines, 
respectively.}
\label{fig:ylepton}
\end{figure}

\begin{figure}
\epsscale{.60}
\plotone{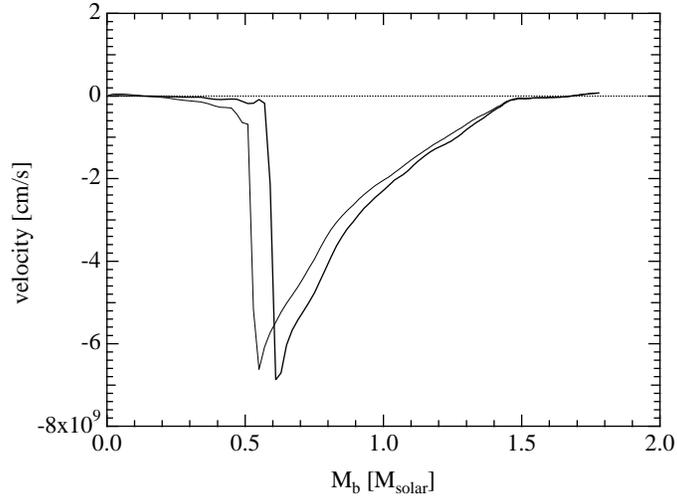}
\caption{Velocity profiles at bounce are shown as a function 
of baryon mass coordinate.
The results for models SH and LS are shown by thick and thin lines, 
respectively.}
\label{fig:velocity}
\end{figure}

The profiles of lepton fractions at bounce are shown in 
Fig. \ref{fig:ylepton}.  
The central electron fraction in model SH is Y$_e$=0.31, 
which is slightly higher than Y$_e$=0.29 in model LS.  
The central lepton fractions including neutrinos for models SH and LS are 
rather close to each other, having Y$_{L}$=0.36 and 0.35, respectively.  
The difference of lepton fraction results 
in a different size of the inner core.
The larger lepton fraction in model SH leads to a larger inner core 
0.61M$_\odot$, whereas it is 0.55M$_\odot$ in model LS.  
Here, the inner core is defined by the region inside the position of velocity 
discontinuity, which is the beginning of shock wave.  
Fig. \ref{fig:velocity} shows the velocity profile 
at bounce.  
We define the bounce (t$_{pb}$=0 ms) as the time 
when the central density reaches 
the maximum, which is similar to other definitions 
such as using the peak entropy height.  
The central density reaches 3.4$\times$10$^{14}$ g/cm$^3$ 
and 4.4$\times$10$^{14}$ g/cm$^3$ in models SH and LS, respectively.  
The difference of stiffness in two EOSs leads to 
a lower peak central density in model SH than that in model LS.  
Because of this difference, the radial size of inner core at bounce 
is $\sim$1 km larger for model SH than that for model LS.  

The initial shock energy, which is 
roughly estimated by the gravitational binding energy of 
inner core at bounce, turns out to be not 
drastically different 
because of the increases both in mass and radial size 
of the inner core in model SH.  
Clearer difference appears at later stages where 
the protoneutron star is formed having a central 
density much higher than the nuclear matter density.  
This is one of reasons why we are interested in the late 
phase of supernova core, where the difference of EOS 
appears more clearly and its influence on the supernova 
dynamics could be seen.  

We remark here that the numerical results with LS-EOS 
at bounce are in good agreement with previous simulations 
such as the reference models by \citet{lie05}.  
For example, the profiles of model LS shown in Figs. \ref{fig:ylepton} 
and \ref{fig:velocity} accord with the profiles of their model G15.  
The behavior after bounce is also qualitatively 
consistent with the reference models up to 250 milliseconds 
(see also section \ref{sec:pns}).  

\subsection{Postbounce phase}\label{sec:postbounce}

The postbounce phase is interesting in many aspects, 
especially in clarifying  
the role of EOS in the neutrino heating mechanism 
and the protoneutron star formation.  
As we have seen in section \ref{sec:shock}, the stall 
of shock wave occurs in a similar manner in two models 
and the difference appears in later stage.  
We discuss here the similarities and the differences
in terms of the effect of EOS.

The evolution of shock wave after it stalls 
around 100 kilometers is controlled mainly 
by the neutrino heating behind the shock wave.
The neutrinos emitted from the neutrinosphere in the 
nascent protoneutron star contribute to the 
heating of material just behind the shock wave 
through absorption on nucleons.  
Whether the shock wave revives or not depends on 
the total amount of heating, hence more specifically, 
on the neutrino spectrum, luminosity, amount of 
targets (nucleons), mass of heating region and 
duration time.  

\begin{figure}
\epsscale{.60}
\plotone{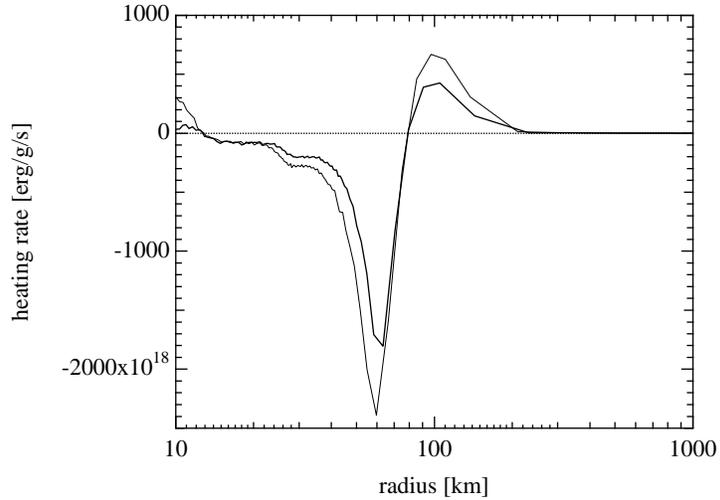}
\caption{Heating rates at t$_{pb}$=150 ms in two models are shown 
as a function of radius.  Notation is the same as in Fig. \ref{fig:velocity}.}
\label{fig:heating}
\end{figure}

\begin{figure}
\epsscale{.60}
\plotone{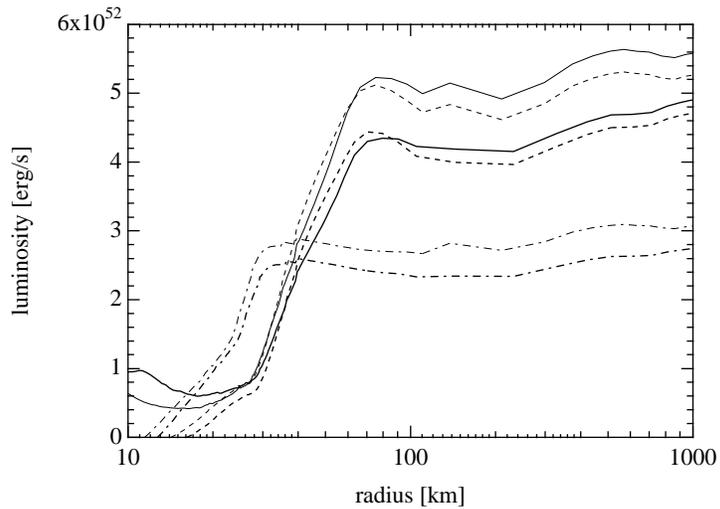}
\caption{Luminosities of $\nu_e$, $\bar{\nu}_e$ and $\nu_{\mu/\tau}$ 
around the heating region are shown by solid, dashed and dot-dashed lines, 
respectively, as a function of radius at t$_{pb}$=150 ms.  
The results for models SH and LS are shown by thick and thin lines, respectively.}
\label{fig:luminositypb}
\end{figure}

\begin{figure}
\epsscale{.60}
\plotone{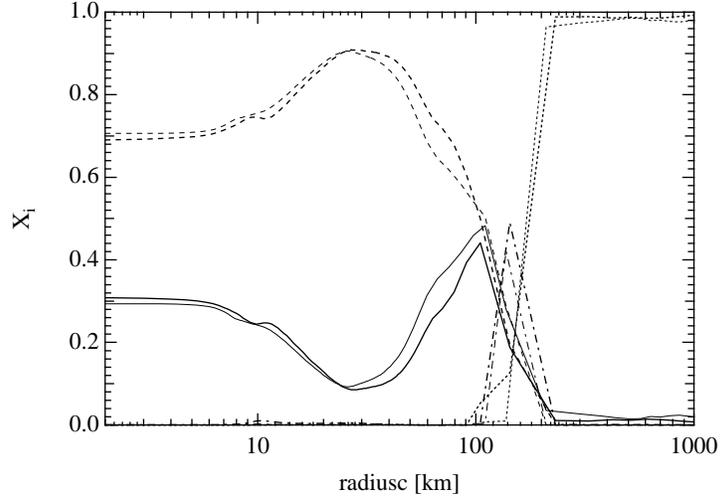}
\caption{Mass fractions in dense matter around the heating region 
are shown as a function of radius at t$_{pb}$=150 ms.  
Notation is the same as in Fig. \ref{fig:xi}.}
\label{fig:xipb}
\end{figure}

The heating rates of material in supernova core 
in two models at t$_{pb}$=150 ms 
are shown in Fig. \ref{fig:heating} as a function of radius.  
The heating rate in model SH is smaller than that 
in model LS around 100 kilometers.  
The cooling rate (negative value in the heating rate) 
in model SH is also smaller than that in model LS.  
The smaller heating (cooling) rate in model SH is 
caused by lower neutrino luminosities and smaller 
free-proton fractions.  
Figs. \ref{fig:luminositypb} and \ref{fig:xipb}
show the radial profiles of neutrino luminosities 
and mass fractions of dense matter 
around the heating region.  
The luminosities in model SH are lower than those in model LS 
for all neutrino flavors.  
The mass fraction of free protons, which are 
the primary target of neutrino heating, is slightly 
smaller in model SH around the heating region.  
These two combinations lower the heating rate 
in model SH.  
It is also interesting that other compositions 
(alpha and nuclei) appear different in this region.

\begin{figure}
\epsscale{.60}
\plotone{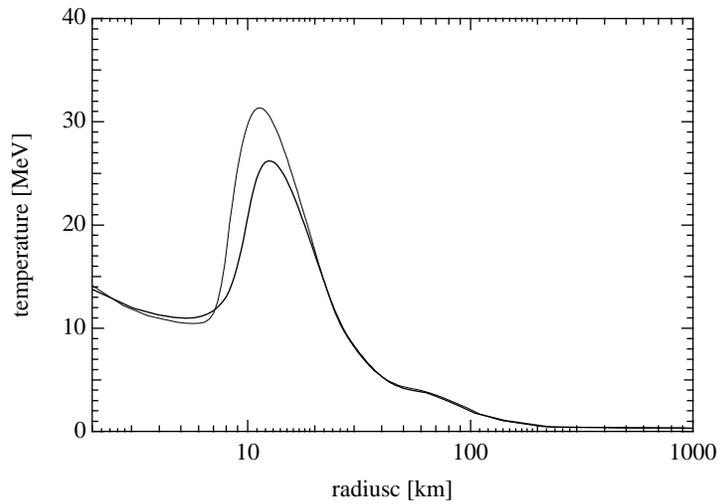}
\caption{Temperature profiles at t$_{pb}$=150 ms for two models are shown 
as a function of radius.  Notation is the same as in Fig. \ref{fig:velocity}.}
\label{fig:temppb}
\end{figure}

The lower luminosities in model SH are related with the 
lower cooling rate.  
The temperature of protoneutron star in model SH is 
generally lower than that in model LS as shown 
in Fig. \ref{fig:temppb}.  
The peak temperature, which is produced by the 
shock heating and the contraction of core, 
in model SH is lower than that in model LS.  
This difference exists also in the surface region 
of the protoneutron star, 
where neutrinos are emitted via cooling processes.  
The temperature at the neutrinosphere in model SH 
is lower and, as a result, the cooling rate is smaller.  
The difference of temperature becomes more evident 
as the protoneutron star evolves as we will see 
in the next section.  

\subsection{Protoneutron star}\label{sec:pns}

\begin{figure}
\epsscale{1.20}
\plottwo{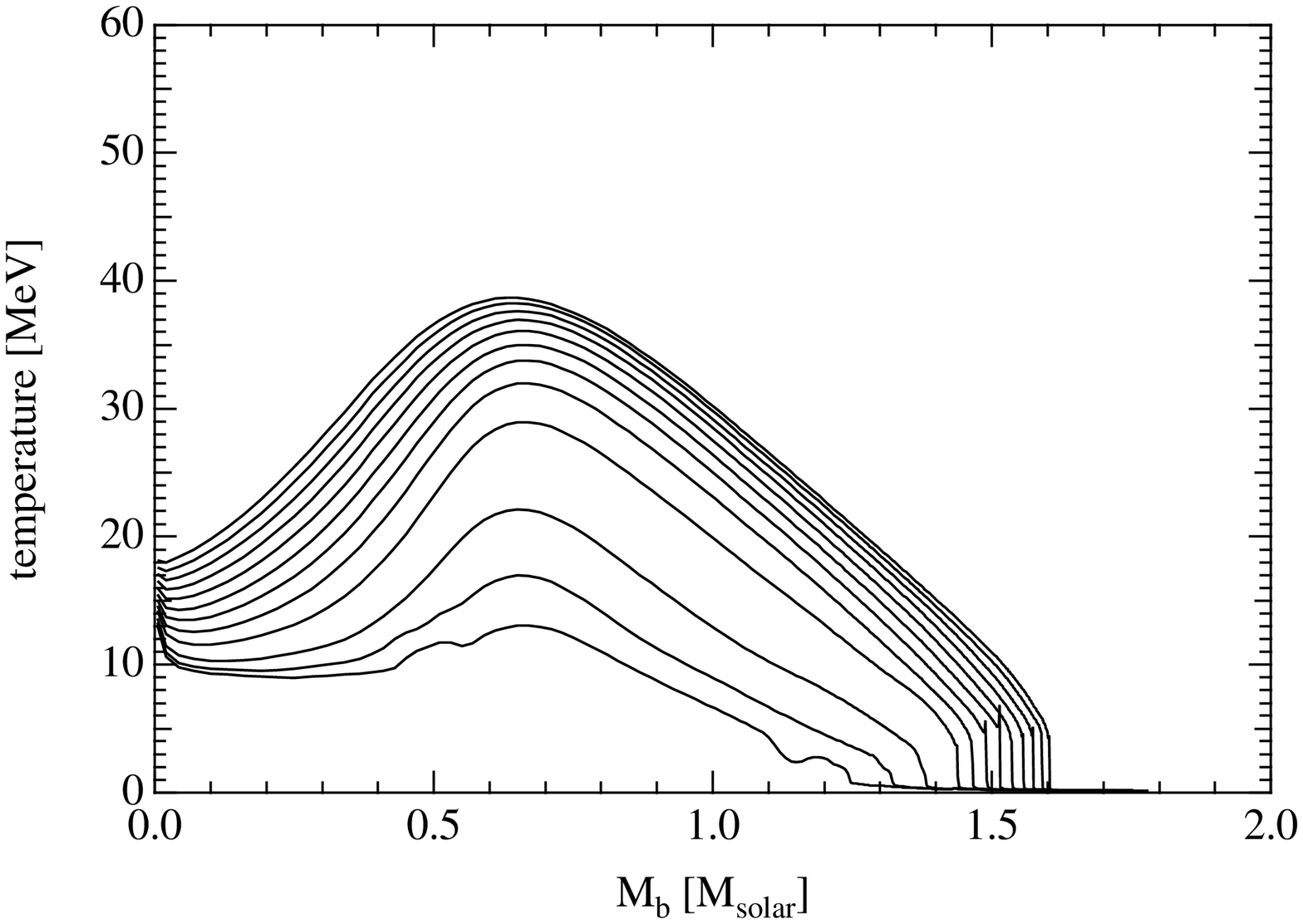}{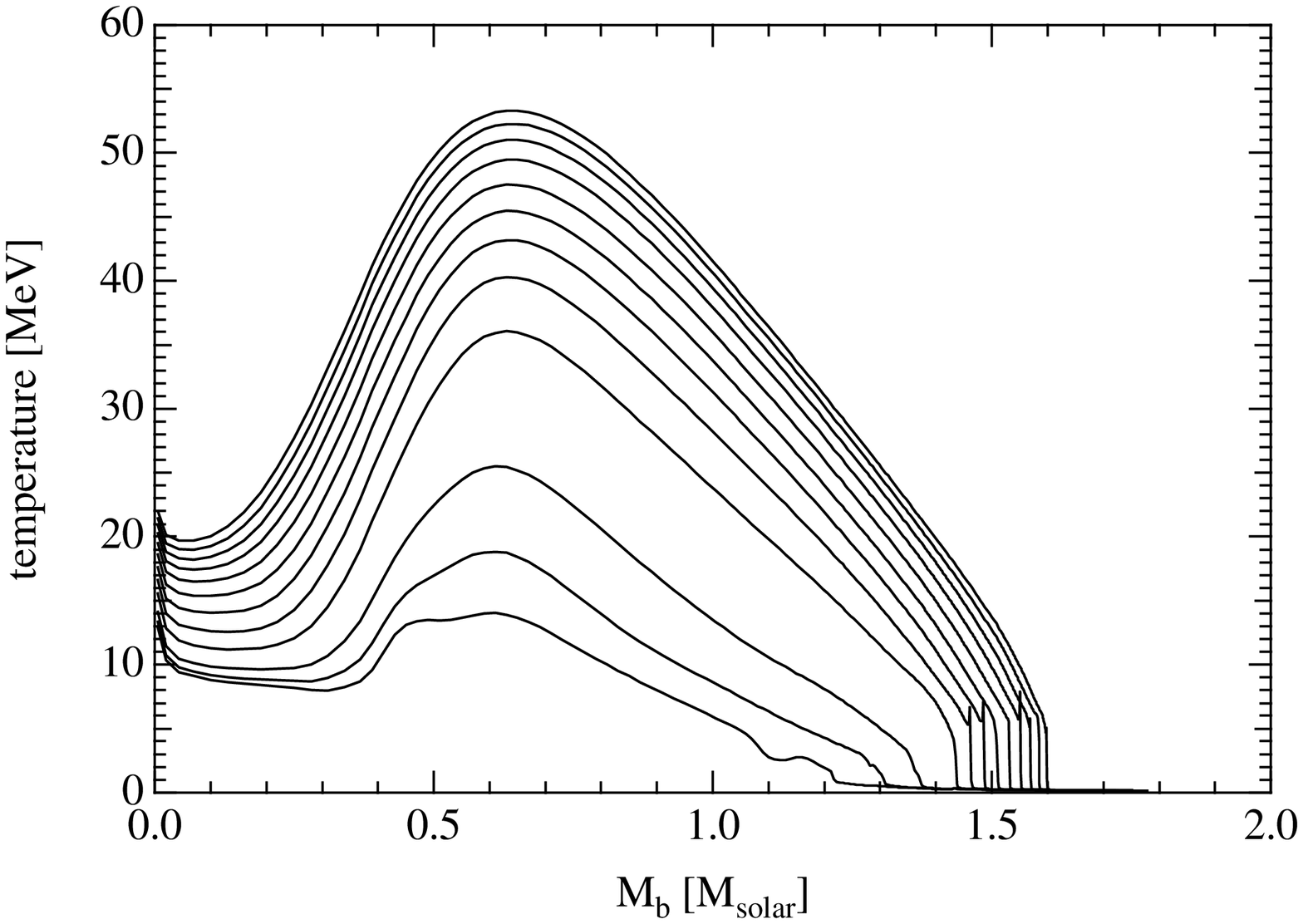}
\caption{Snapshots of temperature profiles as a function of baryon mass 
coordinate from t$_{pb}$=20 ms to t$_{pb}$=1000 ms 
in models SH (left) and LS (right).
Note that small peaks around the central grid are artificial 
due to the numerical treatment.}
\label{fig:pnstemp}
\end{figure}

The thermal evolution of protoneutron star formed 
after bounce is shown in Fig. \ref{fig:pnstemp} 
for two models.  
Snapshots of temperature profile at t$_{pb}$=20, 50, 
100, 200, 300, 400, 500, 600, 700, 800, 900 ms and 1 s are shown.  
The temperature increase is slower in model SH 
than in model LS.  
The peak temperature at t$_{pb}$=1 s is 
39 and 53 MeV in models SH and LS, respectively.  
The temperature difference arises mainly from the stiffness of EOS.  
The protoneutron star contracts more in model LS 
and has a higher central density than in model SH.  
At t$_{pb}$=1 s, 
the central density in model SH is 4.1 $\times$ 10$^{14}$ 
g/cm$^3$ whereas that in model LS is 7.0 $\times$ 10$^{14}$ 
g/cm$^3$, which means the rapid contraction in model LS.  
Since the profile of entropy per baryon is similar 
to each other, lower density results in lower temperature.  
The rapid contraction also gives rise to the rapid 
recession of shock wave down to 20 kilometers in 
model LS.  

We note here on the effective mass.  
In SH-EOS, the effective mass of nucleons is 
obtained from the attraction 
by scalar mesons in the nuclear many-body framework.  
The effective mass at center is reduced to be 440 MeV 
at t$_{pb}$=1 s.  
The nucleon mass is fixed to be the free nucleon mass 
in LS-EOS, on the other hand.  

The temperature difference within 1 second as we have found 
may affect the following evolution of protoneutron star 
up to several tens of seconds, during which the main 
part of supernova neutrinos is emitted.  
Although our models do not give a successful explosion, 
the obtained profiles will still give a good approximation to 
the initial setup for the subsequent protoneutron star cooling.  
Since we have followed the continuous evolution of 
the central core from the onset of gravitational collapse, 
the calculated protoneutron star contains the 
history of matter and neutrinos during the prior stages.  
This is much better than the situation 
so far for calculations of protoneutron star cooling, 
where the profiles from other supernova simulations were adopted 
for the initial model.  
It would be interesting to study the cooling of 
protoneutron star for the two models obtained here.  
Even if such evolutions of protoneutron star are not 
associated with a successful supernova explosion, it 
will be still interesting for the collapsar scenario of GRB and/or 
black hole formation.  
Exploratory studies on various scenarios for the fate of 
compact objects with continuous accretion of matter are 
fascinating and currently under way, 
but it is beyond the scope of the present study.  

\begin{figure}
\epsscale{1.20}
\plottwo{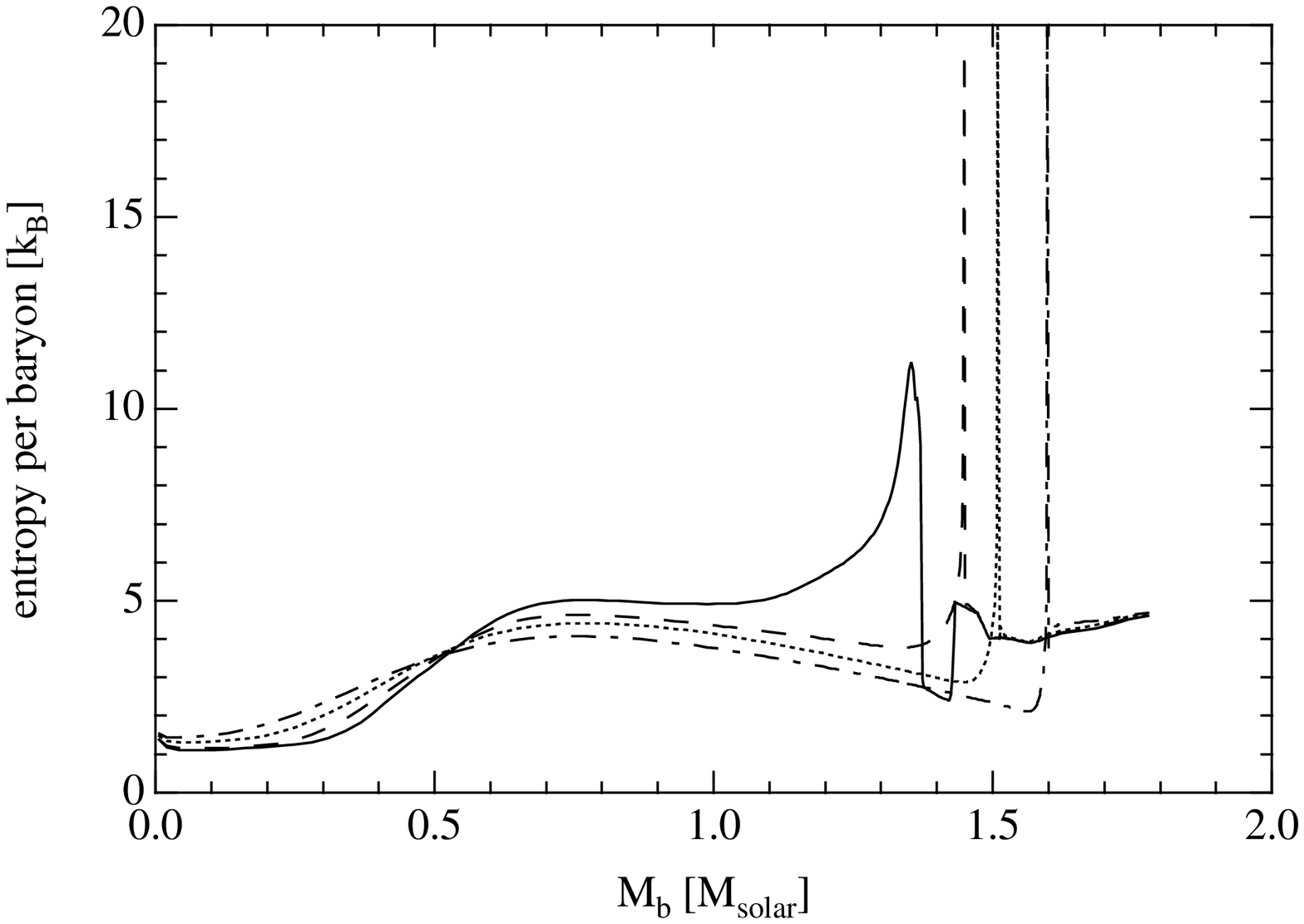}{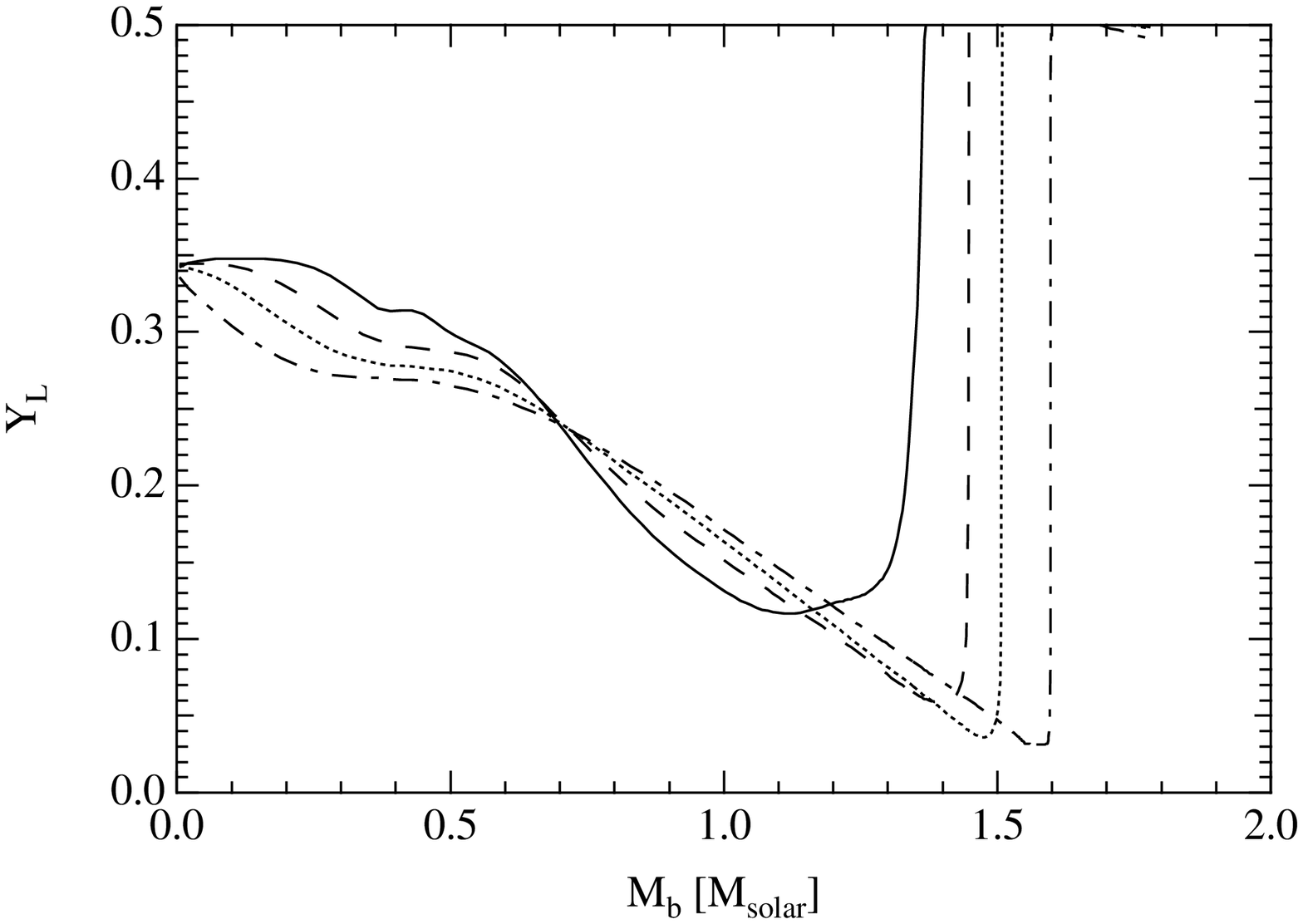}
\caption{Snapshots of entropy (left) and lepton fraction (right) profiles in model LS 
are shown as a function of baryon mass coordinate at t$_{pb}$=100 ms (solid),
t$_{pb}$=250 ms (dashed), t$_{pb}$=500 ms (dotted) and t$_{pb}$=1000 ms (dot-dashed).}
\label{fig:SYlep}
\end{figure}

In Fig. \ref{fig:SYlep}, we display the profiles of entropy 
and lepton fraction in model LS at t$_{pb}$=100, 250, 500 ms and 1 s.  
The distributions of entropy as well as 
other quantities (not shown here) at t$_{pb}$=100 and 250 ms 
are consistent with the reference model G15 \citep{lie05}.  
We have found that the negative gradients in the profiles of 
entropy and lepton fraction commonly appear in late phase 
for both models.  
As for entropy per baryon, the negative gradient appears 
after t$_{pb}$=100 ms in the region between $\sim$0.7M$_\odot$ 
and the shock.  
The negative gradient of lepton fraction appears first 
in the outer core behind the shock 
and then prevails toward the center till t$_{pb}$=1 s.  
Since these regions are unstable against the convection according to 
the Ledoux criterion, the whole region of proto-neutron star 
may be convective after core bounce.  
It has been pointed out that the sign of derivative of thermodynamical 
quantities ($\partial \rho/\partial Y_{\it L}\vert_{P,S}$) 
changes in the neutron-rich environment at high densities 
beyond 10$^{14}$ g/cm$^3$ \citep{sum04}, and the central core may be 
stabilized in model SH.  
Whether the convection occurs efficiently enough to help 
the neutrino-driven mechanism for explosion remains to be 
studied in multi-dimensional $\nu$-radiation-hydrodynamics 
simulations with SH-EOS.  

\subsection{Supernova neutrinos}\label{sec:snnu}

\begin{figure}
\epsscale{1}
\plotone{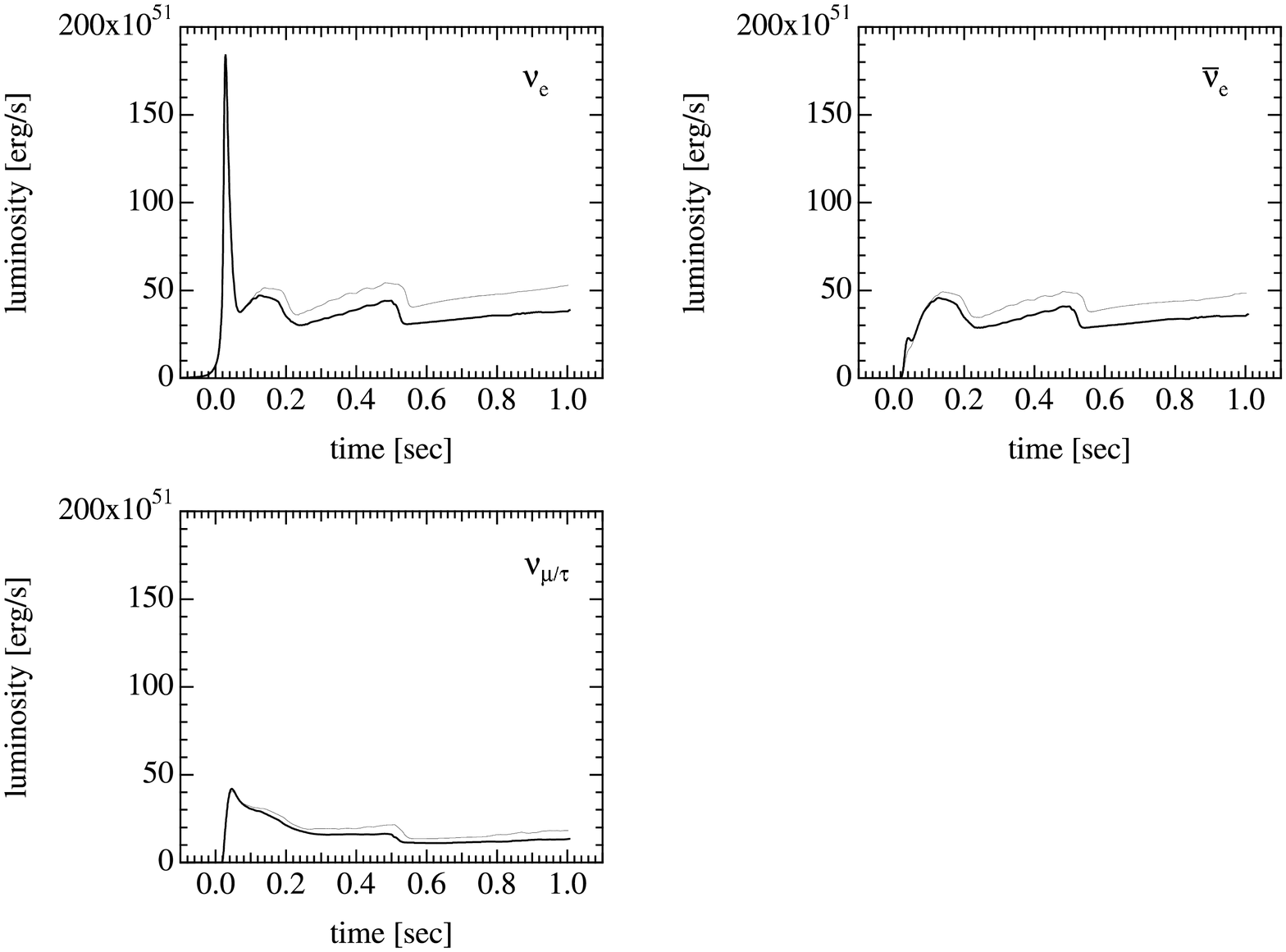}
\caption{Luminosities of $\nu_e$, $\bar{\nu}_e$ and $\nu_{\mu/\tau}$ 
are shown as a function of time after bounce.
Notation is the same as in Fig. \ref{fig:velocity}.  
Kinks around t$_{pb}$=500 ms are 
due to numerical artifact due to the rezoning of mass coordinate.
See the main text for details.}
\label{fig:Lnu}
\end{figure}

The different temperature distribution could affect 
the neutrino luminosities and spectra.  
We discuss here the properties of neutrinos emitted 
during the evolution of supernova core up to 1 second.  
As we have already discussed in section \ref{sec:postbounce}, 
the luminosity of neutrinos in model SH is lower than that 
in model LS after bounce.  
This difference actually appears after t$_{pb}$=100 ms 
as shown in Fig. \ref{fig:Lnu}.  
The initial rise and peak of luminosities in two 
models are quite similar to each other.  
The peak heights of neutronization burst of electron-type 
neutrino are also similar.  
The difference, however, gradually becomes larger and apparent 
after t$_{pb}$=200 ms.  
We remark here that the kinks around t$_{pb}$=500 ms are 
numerical artifact due to the rezoning of mass coordinate 
as discussed in section \ref{sec:rezoning}.  
Except for this kink, the luminosities increase in time.  
For last 150 milliseconds, luminosities show oscillations 
numerically, therefore, we have plotted smoothed curves 
by taking average values.  
It is to be noted that we are interested in the relative 
differences of supernova neutrinos between two models.  

\begin{figure}
\epsscale{1}
\plotone{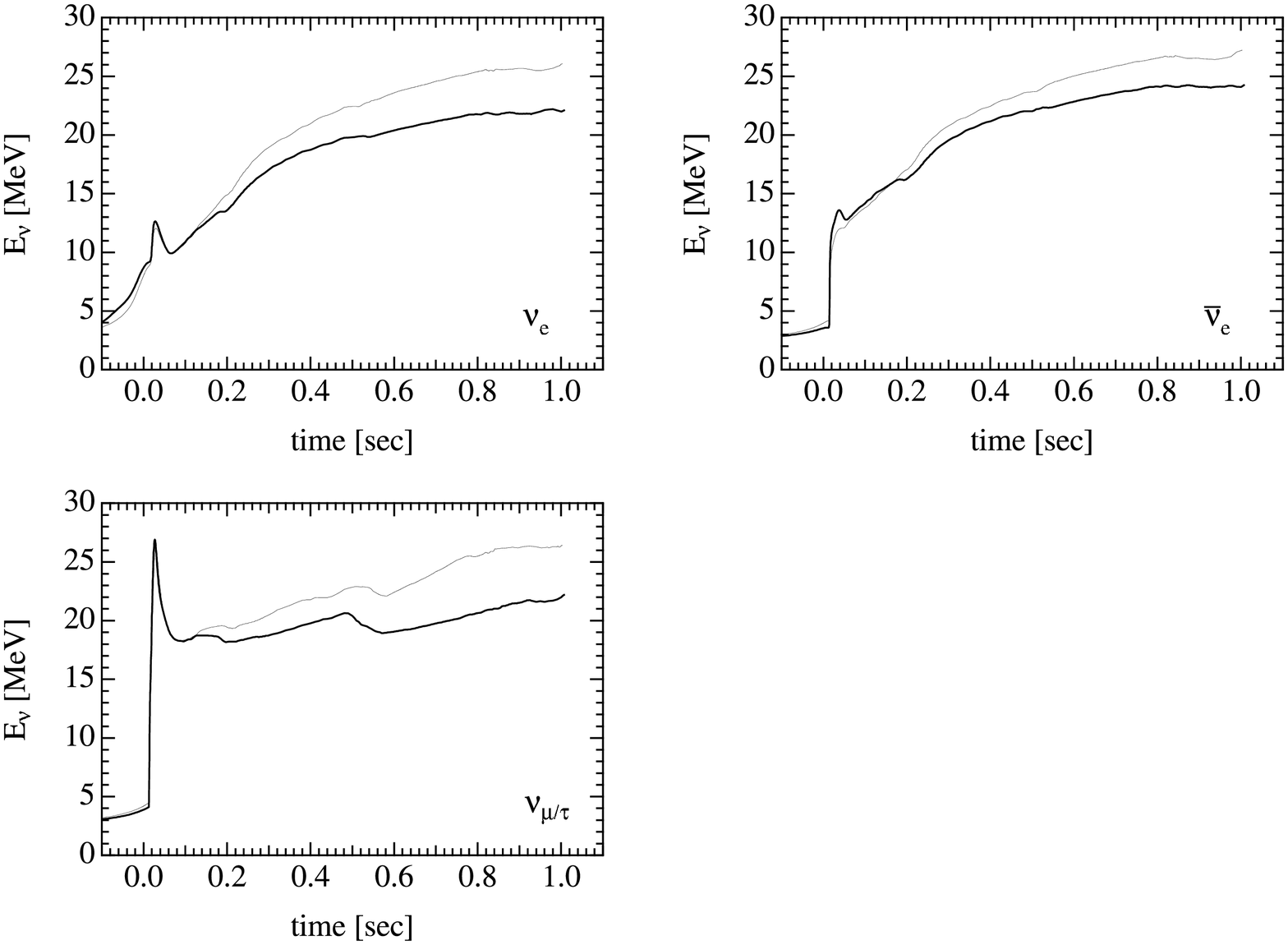}
\caption{Average energies of $\nu_e$, $\bar{\nu}_e$ and $\nu_{\mu/\tau}$ 
are shown as a function of time after bounce.
Notation is the same as in Fig. \ref{fig:Lnu}.
Kinks around t$_{pb}$=500 ms are 
due to numerical artifact due to the rezoning of mass coordinate.
See the main text for details.}
\label{fig:Enu}
\end{figure}

The difference in average energies of neutrinos appears 
in a similar manner to that in luminosities 
as seen in Fig. \ref{fig:Enu}.  
The average energy presented here is the {\it rms} average 
energy, $E_{\nu}=\sqrt{\langle \varepsilon_\nu^2 \rangle}$, 
at the outermost grid point ($\sim$7000 km).  
The average energies up to t$_{pb}$=100 ms are almost identical 
in two models and become different from each other afterwards.  
The average energies in model SH turn out to be lower than 
those in model LS.  
Kinks around t$_{pb}$=500 ms appear due to the same reason 
mentioned above and the curves are smoothed around kinks and 
t$_{pb}\sim$1 s 
to avoid artificial transient behaviors due to the rezoning.  
At t$_{pb}$=1 s, the gap amounts to be more than a few 
MeV and has tendency to increase in time.  
The lower luminosity and average energy in model SH 
is due to the slow contraction of protoneutron star 
and, as a result, the slow rise of temperature as seen 
in Fig. \ref{fig:pnstemp}.  
Again, it would be interesting to see the subsequent cooling phase 
of protoneutron star up to $\sim$20 seconds to obtain 
the main part of supernova neutrinos.  

\section{Summary}

We have performed the numerical simulations of core-collapse supernovae 
by solving general relativistic $\nu$-radiation-hydrodynamics 
in spherical symmetry.  
We have adopted the relativistic EOS table which is based 
on the recent advancement of nuclear many-body theory as well as 
the recent experimental data of unstable nuclei, 
in addition to the conventional Lattimer-Swesty EOS.  
We have done the long-term simulations from the onset of gravitational 
collapse to the late phase far beyond 300 milliseconds after bounce, 
which have not been well studied in previous studies due to the numerical 
restrictions. 
This is meant to explore the chance of shock revival and 
the influence of the new EOS in this stage, and is first such an attempt.  

We have found that a successful explosion of supernova core 
does not occur in neither a prompt nor a delayed way, 
even though we have followed the postbounce evolution up to 1 second 
with the new EOS table.  
The numerical simulation using the Lattimer-Swesty EOS shows 
no explosion either, which is in accord with other recent studies
and in contrast to the finding by Wilson.  
Note that Wilson incorporated convective effects into their spherical 
simulations to obtain successful explosions.  
The shock wave stalls around 100 milliseconds after bounce and recedes down to 
several tens of kilometers to form a stationary accretion shock.  

Regardless of the outcome with no explosion 
we have revealed the differences caused by two EOSs in many aspects, 
which might give some hints for the successful explosion.
We have seen the difference in composition of free-protons and nuclei 
at the collapse phase of supernova core in interesting manners.  
The difference in symmetry energy of two EOSs has caused this effect, 
which can change the electron capture rates and the resulting size of 
bounce cores.  
Although the early shock propagations turn out to be similar 
in the current simulations due to the counter effect by the stiffness of EOS 
and the neutrino heating, 
the implementation of up-to-date electron capture rates on nuclei 
is remaining to be done to obtain more quantitatively reliable 
difference of composition during the collapse phase, which 
may then affect the initial shock energy.  

During the postbounce evolution around 100 milliseconds after bounce 
we have seen that the heating rates in two models are different 
due to the different luminosities and compositions predicted by two EOSs.  
Unfortunately, the merit of larger inner core found in the model 
with SH-EOS is mostly canceled by the smaller heating rate, 
and the behaviors 
of shock wave in the early postbounce phase turn out to be similar in two simulations.  
In general, though, different heating rates by spectral change of neutrinos and compositional 
differences due to EOSs might contribute to the revival of shock wave in the 
neutrino-driven mechanism.  

One of the most important facts we have revealed in the comparison is that 
larger difference actually appears from 200 milliseconds after bounce 
when the central core contracts to become a protoneutron star.  
The temperature and density profiles display larger differences 
as the protoneutron star shrinks further.  
It is in this late phase that we are interested to see possible 
influences of EOS for shock dynamics, since the 
central density becomes high enough and 
the difference of EOS becomes more apparent.  
In the current study, we have not found any shock revival in either model.
We have found, however, distinctly different thermal 
evolution of protoneutron stars in two models, and 
the resulting neutrino spectra are clearly different at this stage.  
This difference might have some influence on the 
accretion of matter.  
The following evolution of protoneutron star 
cooling or formation of a black hole or any other exotic objects 
will certainly be affected.  

After all, the current numerical simulations of core-collapse 
supernovae in spherical symmetry have not given successful explosions, 
even with a new EOS or after long-term evolution.  
One might argue that this situation indicates the necessity of breaking 
spherical symmetry, which is also suggested by some observations and 
has been supported by multi-dimensional simulations.  
However, before one goes to the conclusion that the asymmetry is 
essential in the explosion mechanism, one also has to make efforts 
to find missing ingredients in microphysics (such as hyperons in EOS, 
for example) 
in spherically symmetric simulations.  
Moreover, the spherical simulations serve as a 
reliable basis for multi-dimensional computations of 
$\nu$-radiation-hydrodynamics.  
Convection may be somehow taken into account effectively in spherical 
codes as in the stellar evolution codes.  
These extensions of simulations and microphysics are now in progress.  
The extension of the relativistic EOS table by including strangeness 
particles at high densities has been recently made \citep{ish05}
and corresponding neutrino reactions in hyperonic matter are currently being 
implemented in $\nu$-radiation-hydrodynamics.  

\acknowledgments
K. S. expresses thanks to K. Oyamatsu, A. Onishi, K. Kotake, 
T. Kajino, Tony Mezzacappa and Thomas Janka 
for stimulating discussions and useful suggestions.  
K. S. thanks partial supports from MPA in Garching 
and INT in Seattle where a part of this work has been 
done.  
The numerical simulations have been performed on the 
supercomputers at RIKEN, 
KEK (KEK Supercomputer Project No. 108), JAERI (VPP5000) 
and NAO (VPP5000 System Projects yks86c, rks07b, rks52a).  
This work is supported by the Grant-in Aid for 
Scientific Research (14039210, 14079202, 14740166, 15540243, 15740160)
of the Ministry of Education, Science, Sports and Culture of Japan.
This work is partially supported by 
the Grant-in-Aid for the 21st century COE program "Holistic Research and
Education Center for Physics of Self-organizing Systems".  

\clearpage



\end{document}